\documentclass[usenatbib]{mn2e}
\usepackage{graphicx}
\usepackage{amsmath}
\title[Coevolution of Supermassive Black Holes and their Host Spheroids]{Empirical Constraints on the Coevolution of Supermassive Black Holes and their Host Spheroids}

\author[Li, Conroy and Loeb]{Gongjie Li, Charlie Conroy, Abraham Loeb \\ Harvard-Smithsonian Center for Astrophysics, Cambridge, MA, USA}

\begin{document}
\topmargin-1cm
\bibliographystyle{mn2e}
\maketitle

\newcommand{\apj}{ApJ}
\newcommand{\apjl}{ApJL}
\newcommand{\apjs}{ApJS}
\newcommand{\mnras}{MNRAS}
\newcommand{\aap}{AAP}
\newcommand{\prd}{PRD}
\newcommand{\aj}{AJ}
\newcommand{\pasp}{PASP}
\newcommand{\araa}{ARA\&A}
\newcommand{\nat}{Nature}
\newcommand{\be}{\begin{equation}}
\newcommand{\ee}{\end{equation}}
\newcommand{\bea}{\begin{eqnarray}}
\newcommand{\eea}{\end{eqnarray}}

\def\Mpc{\rm Mpc}
\def\Mbh{M_{\rm BH}}

\def\Msun{M_{\rm \odot}}
\def\kpc{\rm kpc}
\newcommand{\comment}[1]{}



\begin{abstract}

  We investigate the evolution of the $\Mbh-\sigma$ relation by examining the relationship between the intrinsic scatter in the $\Mbh-\sigma$ relation and galaxy bolometric nuclear luminosity, the latter being a probe of the accretion rate of the black hole (BH).  Our sample is composed of galaxies with classical bulges when possible, of which 38 have dynamically measured BHs masses, and 17 have BHs masses measured by reverberation mapping. In order to obtain the bolometric nuclear luminosity for galaxies with low nuclear luminosity, we convert the X-ray nuclear luminosity measured by \textit{Chandra} to bolometric luminosity. We find that the scatter in the $\Mbh-\sigma$ relation is uncorrelated with nuclear luminosity over seven orders of magnitude in luminosity, with the high luminosity end approaching the Eddington luminosity, $\sim L_{Edd}$. This suggests that at the present epoch galaxies evolve {\it along} the $\Mbh-\sigma$ relation.  This conclusion is consistent with the standard paradigm that BHs grow contemporaneously with their host stellar spheroids.

\end{abstract}



\section{Introduction}
\label{s:intro}

The existence of a supermassive BH in the centers of galaxies is well-established \citep{Kormendy95, Richstone, Kormendy01}. Strong links exist between the supermassive BH and host galaxy properties, as evidenced by the BH mass-stellar velocity dispersion ($\Mbh-\sigma$) relation \citep{Ferrarese, Gebhardt}, amongst others. The $\Mbh-\sigma$ relation has been demonstrated not to be a selection effect by \citet{Gultekin11}, to be the strongest correlation between $\Mbh$ and galaxy properties by \citet{Beifiori11}, and has been estimated most recently by \citet{Gultekin} based on 49 $\Mbh$ measurements and 19 $\Mbh$ upper limits, by \citet{Graham} based on 64 $\Mbh$ measurements, and by \citet{Beifiori11} based on 49 $\Mbh$ measurements and 94 $\Mbh$ upper limits.

Despite the wealth of observational data, the origin of this relation is not firmly established. There are several theoretical models that explain the origin of the observed correlations between BH mass and galaxy properties. Feedback from BH accretion on the hosting galaxy is one proposal \citep{Silk98, Fabian, Burkert01}. Simulations often involve galaxy mergers with strong inflow of gas that feeds the BH, powers the quasar and expels enough gas to quench both star formation and further fueling of the BH \citep{Kauffmann, Wyithe03, DiMatteo, Murray05, Robertson06, Johansson09, Hopkins09, Silk10}. This feedback from active galactic nuclei (AGN) regulates the BH-galaxy systems, and leads to tight BH mass-galaxy property relations.  This scenario predicts that $\Mbh$ and $\sigma$ {\it coevolve}.

Other mechanisms have been proposed to explain the tight correlation between BH mass and galaxy properties.  \citet{Peng07} and \citet{Hirschmann10} construct a model for the origin of the $\Mbh-M_{bulge}$ relation in which mergers do not lead to accretion-based growth on the BH. In this model a tight $\Mbh-M_{bulge}$ relation is established through the central limit theorem.  Recently, \citet{Jahnke11} build on this model by including BH accretion and star formation (these processes are un-correlated in the model). They conclude that a causal link between galaxy growth and BH growth is not necessary for obtaining the observed BH mass-galaxy property relations.

In this article we investigate how galaxies evolve in the $\Mbh-\sigma$ plane, and thereby place constraints on these and other models for the origin and evolution of the $\Mbh-\sigma$ relation.  We focus on local galaxies with classical bulges, and investigate the scatter in the $\Mbh-\sigma$ relation as a function of galaxy nuclear luminosity. Our results indicate a scenario where BH mass and $\sigma$ evolve along the $\Mbh-\sigma$ relation, thereby favoring models where BH mass and host stellar spheroids coevolve.

\section{Method}
\label{s:data}

First, we describe the criteria adopted to select our sample and the methods used to analyze the data. We consider the samples compiled by \citet{Gultekin} and \citet{Graham}, which include galaxies with dynamically measured BHs and stellar velocity dispersions. In addition, we include galaxies studied by \citet{Greene}, who measured the central BH mass via masers. We refer to the sum of these samples as our dynamically-based sample. We consider a second set of galaxies with BHs estimated via reverberation mapping \citep{Woo}.

We select galaxies with classical bulges when possible. This is done because \citet{Kormendy} observe that $\Mbh$ does not correlate with the properties of galaxy disks or pseudobulges, and \citet{Sani} find smaller intrinsic scatter of BH mass-host galaxy property relations when excluding galaxies with pseudobulges.  According to bulge classification of the galaxies included in \citet{Greene}, whose classification relies on the galaxy morphology and stellar population property, we select the only galaxy (N1194) that has a classical bulge with nuclear luminosity at $10^{-2.17}~L_{Edd}$, and we also select the galaxy (UGC 3789), which has an unclassified bulge with nuclear luminosity at $ 10^{-0.82}~L_{Edd}$. For the other dynamically-based galaxies, we select classical bulge galaxies according to \citep{Sani}, who classify galaxies with classical bulges by selecting galaxies which have S\'{e}rsic indices higher than two. We include all the reverberation-based galaxies as it is difficult to classify the morphology of galaxies with AGN.

Next, we estimate the bolometric nuclear luminosity. First we consider the sample of \citet{Greene}, who estimate the nuclear bolometric luminosity for their sample using O[III], which is strong and ubiquitous in obscured megamaser systems \citep{Kauffmann03, Zakamska03}. In this approach, O[III] luminosity is converted to $M_{2500}$, where $M_{2500}$ is the magnitude at 2500{~\AA}, and then to bolometric luminosity following the $M_{2500}-L$[OIII] \citep{Reyes08} and $M_{2500}$ bolometric correction \citep{Richards06} for unobscured quasars. The total uncertainty introduced is $\sim 0.5$ dex \citep{Liu09}. This is smaller than our smallest bin size (one dex) when we compare the scatter in the $\Mbh-\sigma$ relation with respect to nuclei luminosity.

In order to obtain the nuclear luminosity for the \citet{Gultekin} and \citet{Graham} samples, we select galaxies with nuclear luminosity measured in the soft X-Ray band by the \textit{Chandra} X-ray observatory \citep{Pellegrini10, Pellegrini05, Zhang, Gonz}. At the lowest luminosity in our sample ($\sim2\times10^{38}{\rm~erg~s^{-1}}$), the central X-ray source has luminosity comparable to the X-ray binary population($\sim10^{38-40}{\rm~erg~s^{-1}}$) \citep{King01}. The use of \textit{Chandra} data is therefore essential because it has sufficient angular resolution to isolate galactic nuclear from bright X-ray binaries. For the reverberation-based sample, we select those with known X-ray luminosity in the NASA/IPAC Extraglactic Database (NED)\footnote{\texttt{http://ned.ipac.caltech.edu/}}. Because these galaxies have nuclear luminosity($\sim2\times10^{43}{\rm~erg~s^{-1}}$) much higher than the X-ray binaries, the X-ray luminosity of the galaxy is dominated by the AGN. Thus, we associate the X-ray luminosity of the galaxy with that of the AGN. Most of the X-ray data are obtained from XMM-Newton \citep{Bianchi09, Markowitz09, Nandra07}, except for Mrk202 and IC 120, which are obtained from ASCA \citep{Ueda05} and BeppoSAX \citep{Verrecchia} separately (detailed properties and references see Table 2). In order to obtain the X-ray luminosity from X-ray fluxes, we estimate distances assuming $H_0 = 73\rm~km~s^{-1}Mpc^{-1}$, $\Omega_m = 0.27$, $\Omega_\Lambda = 0.73$ for $z>0.01$ galaxies, with redshift measurements obtained from NED, which compiled multiple consistent redshift measurements for each galaxy. For $z<0.01$ galaxies, we obtain distances from the Extragalactic Distance Database, which gives updated best distances for galaxies within $3000 \rm~km~s^{-1}$  \citep{Tully09}.

To convert the X-ray luminosity to bolometric luminosity, we first convert the X-ray luminosity of different bands in the literature to luminosity in the band $2-10\,\rm{keV}$ assuming an energy index of $-1$ ($\nu f_{\nu} = constant$) with an uncertainty factor of $\sim2$ (Martin Elvis, private communication). Then, we convert the X-ray luminosity to bolometric luminosity by the bolometric correction for AGNs: $L_{bol}/L_X = 15.8$, with an uncertainty of $\sim0.3$ dex  \citep{Ho09}. Therefore, the nuclear bolometric luminosity is calculated as:\\
\noindent
\begin{align}
L_{bol}=15.8L_X\frac{\ln(10/2)}{\ln(E_2/E_1)},
\end{align}
\noindent
where $E_2$ and $E_1$ represent the upper and lower bound of the observed X-ray band.

We include the properties of the BH and host spheroids for our dynamically-based and reverberation-based sample in Table 1 and Table 2 separately.
\begin{table*}
\caption{Properties of Dynamically-based sample}
\begin{tabular}{|l||l|l|l|l|l|l|}
\hline
Name	&	$\sigma$	&	$\epsilon_{\sigma}$	&	$\log\Mbh$	&	$\epsilon_{\log\Mbh}$	 &	 log[$\frac{L_{bol}}{L_{Edd}}$]	& ref \\
\   &  $\rm{km~s^{-1}}$ & $\rm{km~s^{-1}}$ & $\Msun$ & $\Msun$ & \ & \ \\
(1) & (2) & (3) & (4) & (5) & (6) & (7) \\
\hline
IC1459 	(a)	&	340	&	17	&	9.45	&	0.34	&	-5.52	&	(1)	\\
M31 	(a)	&	160	&	8	&	8.18	&	0.09	&	-8.67	&	(5)	\\
M32 	(a)	&	75	&	3	&	6.49	&	0.08	&	-7.48	&	(1)	\\
M60 	(a)	&	385	&	19	&	9.32	&	0.37	&	-8.17	&	(1)	\\
M81 	(a)	&	143	&	7	&	7.90	&	0.06	&	-4.15	&	(4)	\\
M84 	(a)	&	296	&	14	&	9.18	&	0.35	&	-6.62	&	(1)	\\
M87 	(a)	&	375	&	18	&	9.56	&	0.37	&	-5.70	&	(1)	\\
N524 	(b)	&	253	&	25	&	8.92	&	0.10	&	-7.27	&	(1)	\\
N1194	(c)	&	148	&	24	&	7.82	&	0.05	&	-2.17	&	(6)	\\
N2787 	(a)	&	189	&	9	&	7.63	&	0.04	&	-6.35	&	(3)	\\
N2974	(b)	&	227	&	23	&	8.23	&	0.08	&	-4.85	&	(1)	\\
N3115 	(a)	&	230	&	11	&	8.98	&	0.13	&	-7.19	&	(1)	\\
N3227 	(a)	&	133	&	12	&	7.18	&	0.32	&	-5.12	&	(1)	\\
N3245 	(a)	&	205	&	10	&	8.34	&	0.10	&	-4.52	&	(2)	\\
N3377 	(a)	&	145	&	7	&	8.04	&	0.04	&	-6.74	&	(1)	\\
N3379 	(a)	&	206	&	10	&	8.08	&	0.33	&	-6.90	&	(1)	\\
N3384 	(a)	&	143	&	7	&	7.26	&	0.02	&	-6.09	&	(1)	\\
N3414	(b)	&	237	&	24	&	8.40	&	0.07	&	-7.19	&	(1)	\\
N3585 	(a)	&	213	&	10	&	8.53	&	0.08	&	-6.54	&	(1)	\\
N3607 	(a)	&	229	&	11	&	8.08	&	0.14	&	-6.22	&	(1)	\\
N3608 	(a)	&	182	&	9	&	8.32	&	0.14	&	-7.05	&	(1)	\\
N4026 	(a)	&	180	&	9	&	8.32	&	0.08	&	-6.85	&	(1)	\\
N4261 	(a)	&	315	&	15	&	8.74	&	0.09	&	-4.58	&	(1)	\\
N4459 	(a)	&	167	&	8	&	7.87	&	0.39	&	-6.41	&	(1)	\\
N4552	(b)	&	252	&	25	&	8.68	&	0.07	&	-6.37	&	(1)	\\
N4596 	(a)	&	136	&	6	&	7.92	&	0.13	&	-6.40	&	(2)	\\
N4621	(b)	&	225	&	23	&	8.60	&	0.07	&	-6.62	&	(1)	\\
N5128 	(a)	&	150	&	7	&	7.65	&	0.13	&	-3.54	&	(1)	\\
N5813	(b)	&	239	&	24	&	8.85	&	0.07	&	-5.24	&	(1)	\\
N5846 	(b)	&	237	&	9	&	9.04	&	0.04	&	-6.70	&	(1)	\\
N6251 	(a)	&	290	&	14	&	8.78	&	0.18	&	-2.36	&	(2)	\\
N7457 	(a)	&	67	&	3	&	6.61	&	0.25	&	-5.64	&	(1)	\\
UGC3789	(c)	&	107	&	12	&	7.05	&	0.05	&	-0.82	&	(6)	\\
\hline
\end{tabular}
\medskip \\
\begin{flushleft}
\textbf{Notes.} Column 1: galaxy name and references on $\Mbh$, $\sigma$. Column 2: stellar velocity dispersion ($\sigma$). Column 3: measurement error on $\sigma$. Column 4: black hole mass ($\Mbh$). Column 5: measurement error on $\log \Mbh (\Msun)$. Column 6: X-ray luminosity references.\\
\textbf{References.} \\
$\Mbh$ and $\sigma$ measurements (Column 1.)\\
(a) \citet{Gultekin} (b) \citet{Graham} (c) \citet{Greene} \\
Nuclei X-ray luminosity measurements (Column 7.)\\
(1) \citet{Pellegrini10}; (2) \citet{Gonz}; (3) \citet{Pellegrini05}; (4) \citet{Zhang}; (5) \citet{Li11}; (6) \citet{Greene}
\end{flushleft}
\end{table*}

\begin{table*}
\caption{Properties of Reverberation-based sample}
\begin{tabular}{|l||l|l|l|l|l|l|}
\hline
Name	&	$\sigma$	&	$\epsilon_{\sigma}$	&	log$\Mbh$	&	$\epsilon_{log\Mbh}$	 &	 log[$\frac{L_{bol}}{L_{Edd}}$]	& ref \\
\   &  $\rm{km~s^{-1}}$ & $\rm{km~s^{-1}}$ & $\Msun$ & $\Msun$ & \ & \ \\
(1) & (2) & (3) & (4) & (5) & (6) & (7) \\
\hline
3C 120 	&	162	&	20	&	7.72	&	0.23	&	-0.18	&	(5)	\\
Ark 120 	&	221	&	17	&	8.15	&	0.11	&	-0.60	&	(4)	\\
Mrk 79 	&	130	&	12	&	7.70	&	0.16	&	-0.99	&	(6)	\\
Mrk 110 	&	91	&	7	&	7.38	&	0.14	&	0.14	&	(3)	\\
Mrk 202 	&	78	&	3	&	6.13	&	0.22	&	-0.58	&	(2)	\\
Mrk 279 	&	197	&	12	&	7.52	&	0.15	&	-0.52	&	(3)	\\
Mrk 590 	&	189	&	6	&	7.66	&	0.12	&	-0.46	&	(7)	\\
Mrk 1310 	&	84	&	5	&	6.33	&	0.17	&	-0.30	&	(1)	\\
Mrk 1383	&	217	&	15	&	9.09	&	0.16	&	-1.21	&	(3)	\\
N3227 	&	136	&	4	&	7.60	&	0.24	&	-2.06	&	(5)	\\
N3516 	&	181	&	5	&	7.61	&	0.18	&	-1.13	&	(5)	\\
N3783 	&	95	&	10	&	7.45	&	0.13	&	-0.10	&	(3)	\\
N4051 	&	89	&	3	&	6.18	&	0.19	&	-1.05	&	(8)	\\
N4151 	&	97	&	3	&	7.64	&	0.11	&	-1.46	&	(4)	\\
N4253	&	93	&	32	&	6.23	&	0.30	&	0.11	&	(3)	\\
N4593 	&	135	&	6	&	6.97	&	0.14	&	-1.07	&	(4)	\\
N5548 	&	195	&	13	&	7.80	&	0.10	&	-0.99	&	(4)	\\
N6814 	&	95	&	3	&	7.25	&	0.12	&	-1.52	&	(4)	\\
N7469 	&	131	&	5	&	7.06	&	0.11	&	-0.32	&	(9)	\\
\hline
\end{tabular}
\medskip\\
\begin{flushleft}
\textbf{Notes.} Column 1: galaxy name. Column 2: stellar velocity dispersion ($\sigma$). Column 3: measurement error on $\sigma$. Column 4: black hole mass ($\Mbh$). Column 5: measurement error on $\log \Mbh (\Msun)$. Column 6: X-ray luminosity references. \\
\textbf{References.} \\
Nuclei X-ray luminosity measurements and instruments (Column 7.)\\
(1) \citet{Verrecchia}, BeppoSAX; (2a) \citet{Bianchi09}, XMM; (2b) \citet{Markowitz09}, XMM; (2c) \citet{Nandra07}, XMM; (3) \citet{Ueda05}, \emph{ASCA} \\
Local galaxy (z$<$0.01) distance measurements:
N3227, N3516, N3783, N4051, N4151, N4593: \citet{Tully09};
 \end{flushleft}
 \end{table*}

In reality the correction factor, $L_{bol}/L_X$, depends on the nuclear luminosity: low luminosity AGNs tend to be ``X-ray-loud" \citep{Ho99, Ho09}. In other words, the lower X-ray nuclear luminosity corresponds to an even lower bolometric nuclear luminosity and vice versa. Note that this additional complexity does not mix the order of galaxies with respect to their nuclei luminosity. As we compare galactic properties of lower nuclear luminosity galaxy to those of higher nuclear luminosity galaxies, without computing the exact bolometric luminosity, our conclusion is not affected by assuming a constant bolometric correction factor. Because the bolometric correction factor only introduces an uncertainty of $\sim0.3$ dex \citep{Ho09}, the total uncertainty of the bolometric luminosity is less than the smallest bin width, one dex. Therefore, we expect our results to be largely unaffected by the uncertainties in nuclear bolometric luminosity.

In total, we have 38 galaxies with dynamically measured BH mass and 17 galaxies with reverberation-mapping BH mass measurements in our sample. For the dynamically-based sample, the range of the nuclear luminosity is limited because the nuclear luminosity is difficult to measure for low nuclear luminosity galaxies and the BH mass is difficult to estimate dynamically for high nuclear luminosity galaxies. The reverberation mapping measurements are normalized by setting a constant virial coefficient so that the reverberation mapping-based and dynamically-based $\Mbh-\sigma$ relations agree.  The assumption of a constant virial coefficient could potentially introduce a larger scatter for the reverberation-based sample. In addition, the virial coefficient may depend on the nuclear luminosity. If so, choosing a constant virial coefficient may introduce a BH mass uncertainty that depends on the nuclear luminosity. This could affect our result. In order to mitigate this potential bias, we do not rely heavily on comparing galaxies across our two samples.

With $\Mbh$, $\sigma$ and nuclear luminosity in hand, we are now in a position to consider the scatter in the $\Mbh-\sigma$ relation with respect to nuclear luminosity. In order to obtain the scatter, we perform a linear fit to the $\Mbh-\sigma$ relation by minimizing $\chi^2$ for the dynamically-based sample, as done by \citet{Tremaine}. Our best-fit parameters are consistent with those in the literature: $\log(\Mbh / \Msun) = 8.27 + 4.05\log (\sigma/200 \,\rm{km~s^{-1}})$, with an intrinsic scatter of 0.27. The scatter between the measured $\Mbh$ for an individual galaxy and the $\Mbh-\sigma$ relation, in units of the uncertainty, is calculated as:
\noindent
\begin{align}
scatter=\frac{\Delta M_{BH}}{\sqrt{\epsilon_{M}^2+b^2\epsilon_{\sigma}^2}},
\end{align}
\noindent
where $\Delta M_{BH}$ is the difference between the measured BH mass and the BH mass corresponds to the fitted $\Mbh-\sigma$ relation, $\epsilon_{M}$ is the measurement uncertainty of BH mass, $\epsilon_\sigma$ is the measurement uncertainty of the velocity dispersion and b is the slope of the $\Mbh-\sigma$ relation. For a fixed slope of the $\Mbh-\sigma$ relation, the scatter in BH mass determines the scatter in $\sigma$ ($scatter_{BH}=b \times scatter_{\sigma}$). Measuring the scatter in BH mass makes it more convenient to compare with the literature, where intrinsic scatter in BH mass is discussed (eg. \citealt{Tremaine}).

\section{Results}
\label{s:res}

In Figure 1 we plot the $\Mbh-\sigma$ relation for our sample. In order to compare the scatter in the $\Mbh-\sigma$ relation with the nuclear luminosity, we color code the symbols by the nuclear luminosity level, $\log L_{bol}/L_{Edd}$. Because the reverberation mapping method has a higher uncertainty in BH mass (because a constant virial coefficient is adopted), the scatter in the reverberation-based sample is slightly larger than that for the dynamically-based sample. It is apparent already from this figure that there is no strong correlation between $\Mbh-\sigma$ scatter and nuclear luminosity within each sample.

\begin{figure}
\begin{center}
\resizebox{3.5in}{!}{\includegraphics{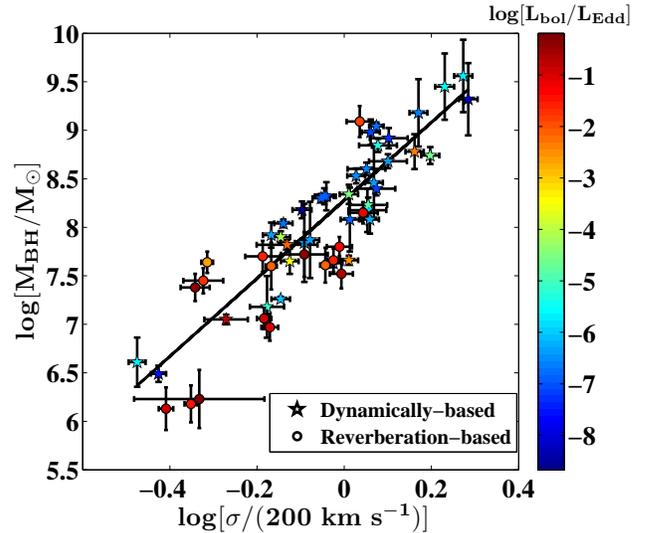}}
\caption{$\Mbh-\sigma$ relation for galaxies with classical bulges and dynamically-based BH masses (stars), and reverberation-based masses (circles). The color bar indicates the nuclear luminosity levels, $\log L_{bol}/L_{Edd}$. This figure demonstrates visually that there is no strong correlation between the scatter and the nuclear luminosity.}
\label{fig:f1}
\vspace{0.1cm}
\end{center}
\end{figure}

In order to investigate the dependence of the scatter on nuclear luminosity quantitatively, we show in Figure 2 the scatter versus the nuclear luminosity (top panel) and the average scatter in bins of the nuclear luminosity level (bottom panel). To keep track of the standard deviation of the scatter at each nuclei luminosity bin, we plot the standard deviation in each bin as vertical dash error bars. We choose bins so that the number of galaxies per bin is comparable, in order to minimize Poisson noise. Specifically, the number of galaxies in each bin are 13, 13, 12, 9, 8. Notice that we do not mix the dynamically-based and reverberation-based samples. The scatter and the standard deviation remain approximately constant as the nuclear luminosity increases. This is the basic result of this article.

\begin{figure}
\begin{center}
\mbox{\includegraphics[width=3.5in]{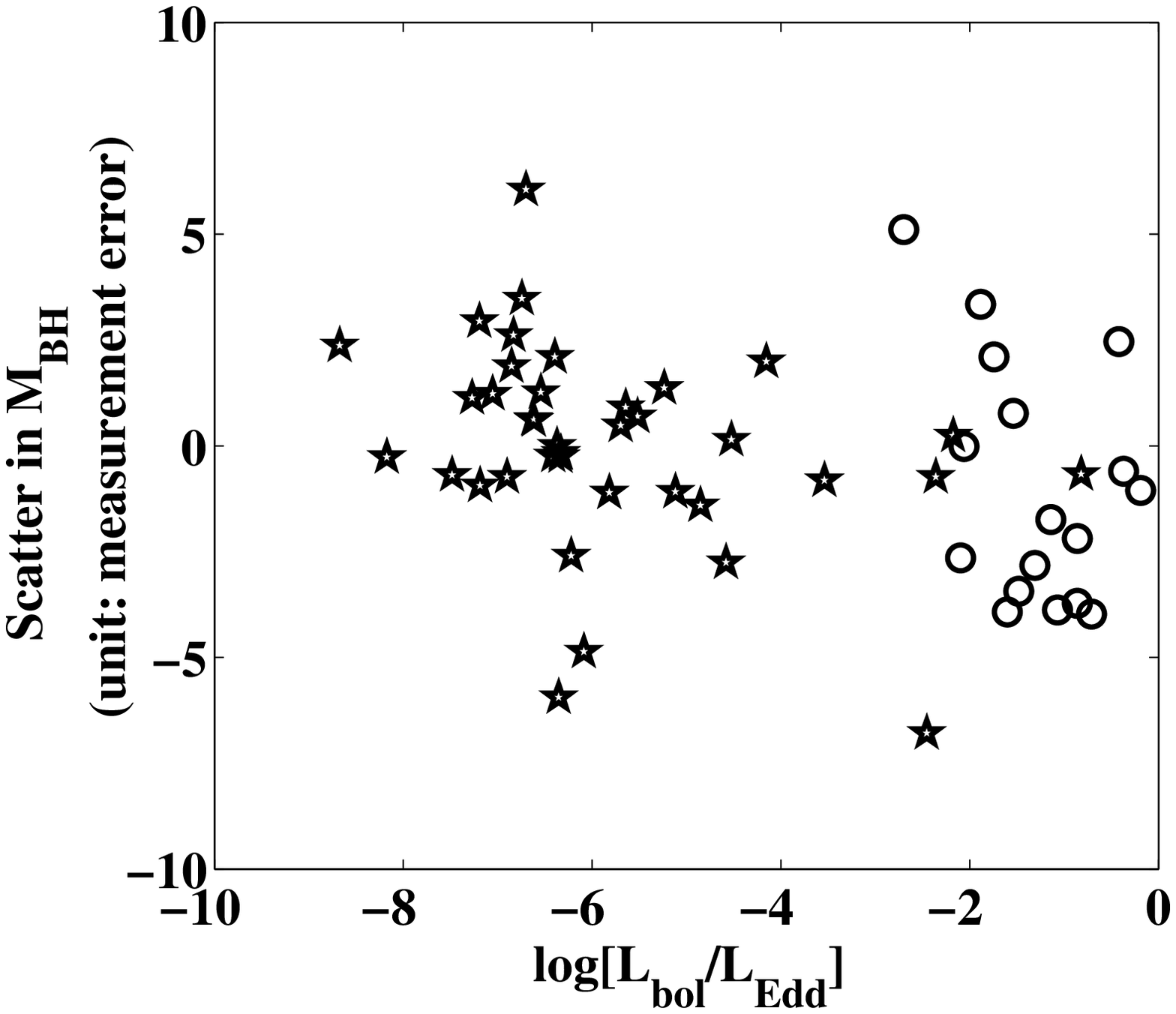}}\\
\mbox{\includegraphics[width=3.5in]{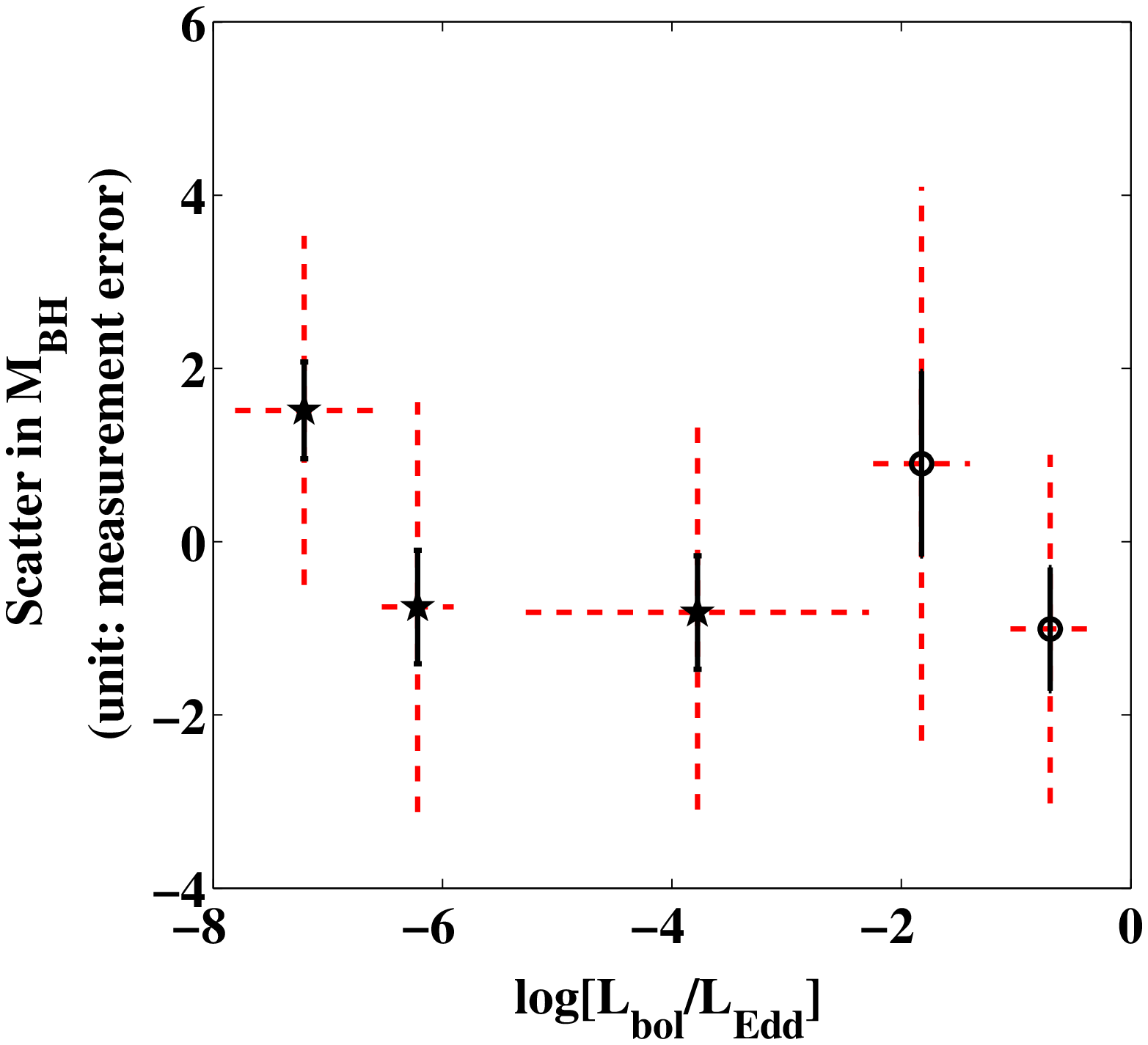}}
\caption{Scatter (top panel) and binned scatter (bottom panel) in the $\Mbh-\sigma$ relation as a function of galaxy nuclear luminosity. Measurement error is taken to be the sum of the measurement uncertainty in $\Mbh$ and $\sigma$, where the measurement uncertainty in $\sigma$ is scaled to $\Mbh$ (see scatter definition in equation (2)). Galaxies with dynamical BH measurements are represented by stars, and those with reverberation measurements are represented by circles. The solid vertical error bars indicate the error on the mean scatter in each bin; the dashed vertical error bars indicate the standard deviation of the scatters in each bin; the horizontal error bars indicate the standard deviation of the nuclear luminosity levels in each bin. It is clear from this figure that the scatter is independent of nuclear luminosity.}
\label{fig:f2}
\vspace{0.1cm}
\end{center}
\end{figure}

This result can be seen another way by considering the timescales that govern the growth of the BH mass and the dynamical time.  We define the BH growth timescale as:
\begin{align}
t_{acc} \nonumber &\equiv \frac{\Mbh}{\dot M_{\rm BH}}\\
&= 4\times10^7 \bigg( \frac{\epsilon}{0.1} \bigg) \bigg(\frac{L_{bol}}{L_{Edd}}\bigg)^{-1} {\rm yr},
\end{align}
where $\epsilon$ is the radiative efficiency, which we take to be 0.1, $L_{Edd} = 3.5\times10^4 (\frac{\Mbh}{M_{\odot}}) L_{\odot}$ is the Eddington luminosity.

The dynamical time is defined via:
\begin{align}
t_{dyn} &\equiv\frac{R_e}{\sigma}.
\end{align}
where $R_e$ is the effective radius of the galaxy. Because we are interested in the two timescales when the BH is accreting rapidly, we set an arbitrary threshold ($4\times10^{-3}L_{Edd}$) and select galaxies with nuclear bolometric luminosity higher than this value. Our result is independent of this threshold. This leaves us with 3 dynamically-based measurements, and all the reverberation-based measurements. Then, to calculate $t_{dyn}$, we obtain the effective radii of these galaxies from the literature \citep{Sani, Bentz, Lauer, Marconi}. For the galaxies with no measured effective radii in the literature (N1194, UGC3789, Mrk 202 and N4253), we use the galaxy radius calculated as the following. We estimate the radii of the galaxy by multiplying the angular radii by the angular-size distances. We obtain the angular radii from the 2MASS isophotal measurements with reference level of the radii set at $20~\rm K-band~magnitude~arcsec^{-2}$. The angular-size distance is calculated assuming $H_0 = 73 \rm~km~s^{-1}Mpc^{-1}$, $\Omega_m = 0.27$, $\Omega_\Lambda = 0.73$ as the four galaxies are all $z>0.01$ galaxies. The redshift of the galaxies are obtained from NED, which compiled multiple consistent redshift measurements for each galaxy.

In Figure 3 we plot the ratio of $t_{acc}$ to $t_{dyn}$ as a function of the nuclear luminosity in units of the Eddington luminosity. We fit a straight line and find the slope of the line is $-0.94$ and the ratio approaches unity when the nuclear luminosity approaches the Eddington limit. As discussed in section 2, the bolometric correction factor depends on the nuclei luminosity. By fixing a constant correction factor, the nuclear bolometric luminosity is probably underestimated at high luminosity by roughly factor of 2. This causes $t_{acc}$ to be overestimated at high luminosity by the same factor. Thus, plotting the ``true'' $t_{acc}$ versus the ``true'' luminosity, which are overestimated and underestimated separately by the same factor, the points in Figure 3 should shift to the left along our observed trend. Therefore, the trend in Figure 3 is robust against the uncertainties in nuclear luminosity.

\begin{figure}
\begin{center}
\resizebox{3.5in}{!}{\includegraphics{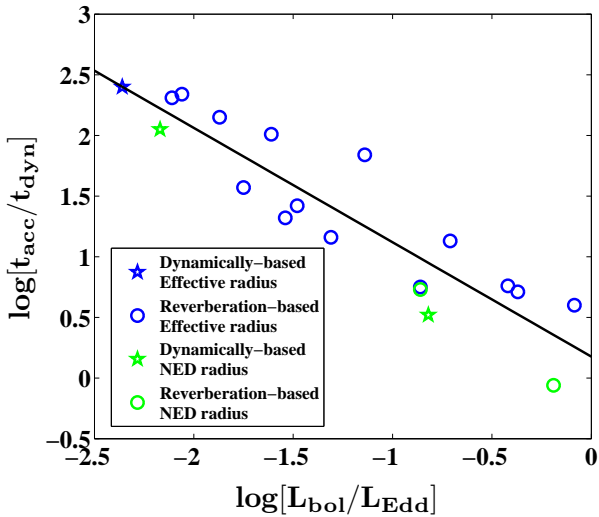}}
\caption{Ratio of the instantaneous accretion timescale to dynamical timescale as a function of bolometric nuclear luminosity in units of the Eddington luminosity. The black line indicates the linear fit of the relation, which has a slope of -0.94. The fact that the ratio of timescales approaches unity as the luminosity approaches Eddington implies that BH mass and $\sigma$ coevolve.}
\label{fig:f3}
\vspace{0.1cm}
\end{center}
\end{figure}

\section{Discussion}
\label{s:dis}

The relationship between scatter in the $\Mbh-\sigma$ relation and the accretion rate of the BH puts interesting constraints on how galaxies evolve in the $\Mbh-\sigma$ plane.  For example, if BH growth and host stellar spheroid growth are uncorrelated then we would expect that galaxies with rapidly accreting BHs to lie systematically off of the $\Mbh-\sigma$ relation.  In contrast, if BHs and spheroid coevolve, then we would expect no correlation between scatter in $\Mbh-\sigma$ and BH activity, which is precisely what we observe.  Our results therefore favor the scenario wherein spheroid and BHs coevolve along the $\Mbh-\sigma$ relation.

In support of our conclusion, we also find that the dynamical time of the galaxy becomes comparable to the BH accretion timescale when the nuclear luminosity approaches the Eddington luminosity (Figure 3). We can understand how galaxies populate the space in Figure 3 in light of our results. Galaxies cannot be in the upper right region of Figure 3 because otherwise the BH would grow much slower than $\sigma$, resulting in a correlation between scatter and nuclear luminosity, which is not observed. We can also understand why there are no galaxies in the lower left region of Figure 3: there is no limit to how long $t_{acc}$ can be (because the accretion rate can be arbitrarily close to zero), while there is a limit to how small $t_{dyn}$ can be.  So $t_{acc}/t_{dyn}$ can increase at lower $L/L_{Edd}$. For systems experiencing rapid growth in the BHs, the growth timescale of the BH is comparable to the dynamical time, supporting the idea that galaxies evolve along the $\Mbh-\sigma$ relation.

As a further extention, we also consider the intrinsic scatter of galaxies with $\Mbh$ measured by the virial method \citep{Xiao, Shen} for high redshift AGN. We find that the intrinsic scatter of those is similar to the intrinsic scatter of the galaxies with BH masses measured by the reverberation-mapping method. The virial method uses the radius-luminosity relation to estimate the radius of the broad-line region, and then uses the reverberation mapping method formalism to measure BH masses. Thus, the mass of the BH has an even bigger uncertainty. Barring these additional uncertainties, this suggests that high-z galaxies may also evolve along $M-\sigma$ relation.

Our result is inconsistent with models that predict a non-causal origin of BH mass-galaxy property relations. As discussed in \citet{Jahnke11}, in a non-causal origin model, the BH mass growth rate is uncorrelated with the growth of $\sigma$, and the BH mass-galaxy property relation converge only through the central limit theorem. Thus, when a given BH is growing its mass at high nuclear luminosity, $\sigma$ does not ``catch up''  until after several merger events. Such a model would therefore predict the scatter in $\Mbh-\sigma$ to be larger for higher nuclear luminosity, in contradiction to our results.

On the other hand, our conclusion is consistent with the scenario emerging from simulations that episodes of major spheroid growth and BH growth occur on similar timescales via mergers. Simulations find that the epoch of rapid BH accretion is limited by AGN feedback to be $\sim 100$ Myr, which is similar to the dynamical time (eg. \citealt{Kauffmann, DiMatteo, Sijacki07, Hopkins09, Hopkins11, Blecha}). This suggests that the BH mass and $\sigma$ change concurrently.  Models that appeal to fueling of the BH by recycled gas \citep[e.g.,][]{Ciotti07} must satisfy our constraint that galaxies evolve along $\Mbh-\sigma$ even during high accretion rate phases.

In addition, our result reinforces the assumptions in various studies. For instance, it shows that BH grows simultaneously with the potential well formation in mergers, which is assumed by \citet{Shankar} to investigate the cosmological evolution of the $\Mbh-\sigma$ relation. The self-regulated growth of supermassive BH is assumed when investigating the link between quasars and the red galaxy population by \citet{Hopkins06}, and when constraining the accretion history of massive BHs by \citet{Volonteri06}. Moreover, it constrains the total timescale of the episodic random accretion model proposed by \citet{Wang} to be similar to $t_{dyn}$.

In our dynamically-based sample we only have one galaxy with nuclear bolometric luminosity higher than $0.1L_{Edd}$, and three galaxies with nuclear bolometric luminosity higher than $0.001L_{Edd}$.  We can expect more stringent constraints on the coevolution of galaxies and BHs as additional maser-based BH masses are obtained for higher nuclear luminosity galaxies.


\section*{Acknowledgments}

We thank Martin Elvis, Pepi Fabbiano, Silvia Pellegrini, Junfeng Wang, Yue Shen, Laura Blecha and Phil Hopkins for helpful conversations. This work was supported in part by NSF grant AST-0907890 and NASA grants NNX08AL43G and NNA09DB30A.

\bibliography{ms_refs}

\end{document}